\documentclass[pra,onecolumn,11pt,amsmath, superscriptaddress,notitlepage,showpacs]{revtex4-1}
\usepackage{amsmath,amsfonts,amssymb,amstext,amscd,amsthm,dsfont,bbm,hyperref,natbib}
\usepackage{physics}
\usepackage{color}
\usepackage{soul,xcolor}
\usepackage{gensymb}
\hypersetup{
    colorlinks = true,
    citecolor=blue,
    linkcolor = red,
    anchorcolor = red,
    citecolor = blue,
    filecolor = red,
    pagecolor = red,
    urlcolor = red,
    }
\usepackage[normalem]{ulem}
\usepackage{graphicx}
\usepackage{bbold}
\usepackage{braket}
\usepackage{placeins}

\newtheorem{Theorem}{Theorem}

\newtheorem{Definition}{Definition}

\begin{document}
\title{Measure of not-completely-positive qubit maps: the general case}
	\author{Vinayak Jagadish}
	\affiliation{Quantum Research Group, School  of Chemistry and Physics,
		University of KwaZulu-Natal, Durban 4001, South Africa}\affiliation{ National
		Institute  for Theoretical  Physics  (NITheP), KwaZulu-Natal,  South
		Africa}
			\author{R. Srikanth}
	\affiliation{Poornaprajna Institute of Scientific Research,
		Bangalore- 560 080, India}
	\author{Francesco Petruccione}
	\affiliation{Quantum Research Group, School  of Chemistry and Physics,
		University of KwaZulu-Natal, Durban 4001, South Africa}\affiliation{ National
		Institute  for Theoretical  Physics  (NITheP), KwaZulu-Natal,  South
		Africa}
	
\begin{abstract} 
We show that the set of not-completely-positive (NCP) maps is unbounded, unless further assumptions are made. This is done by first proposing a reasonable definition of a valid NCP map, which is nontrivial because NCP maps may lack a full positivity domain. The definition is motivated by specific examples. We prove that for valid NCP maps, the eigenvalue spectrum of the corresponding dynamical matrix is not bounded. Based on this, we argue that in general the volume measure of qubit maps, including NCP maps, is not well defined.
\end{abstract}
\maketitle

\section{Introduction}

For a long time,  the study  of open  quantum systems~\cite{petruccione} remained  confined to maps which are completely positive (CP). However, phenomenological studies of spin relaxation showed that complete positivity was not a necessary requirement~\cite{simmons_completely_1981,raggio_remarks_1982,simmons_another_1982}. Based on the concept of linear assignment maps, a better view on the issue of complete positivity was brought forward in the debate between Pechukas and Alicki~\cite{pechukas_reduced_1994,alicki_comment_1995,pechukas_pechukas_1995}.
With  the  understanding  of  the importance  of  initial  correlations~\cite{jordan_dynamics_2004,carteret_dynamics_2008},
non-Markovian dynamics~~\cite{breuer_colloquium:_2016,li_concepts_2017}
and advances in  control of quantum systems, not-completely-positive (NCP)  maps are now  studied ever
more actively~~\cite{Szarek-2008,cuffaro_debate_2013}. It is also now known that NCP maps arise naturally as intermediate maps of quantum non-Markovian processes~\cite{rivas_entanglement_2010}.

Within the space of positive maps, the relative measure for  CP and NCP maps was addressed for Pauli unital channels in~\cite{jagadish_measure_2019}. The present work revisits the question of measure of NCP maps, but by relaxing the requirement that the map has a full positivity domain on the system of interest, i.e., all states of the reduced system produce valid output states. An example of this type is the map corresponding to the partial-trace operation. But, more generally, we can allow maps, namely NCP maps, whose positivity domain is restricted. An example of this type would be the intermediate (NCP) map for the non-Markovian dephasing channel~\cite{shrikant_non-markovian_2018}, for which, at the singularity in the decoherence rate, only a set of states of zero measure produces a valid output.

In~\cite{jagadish_measure_2019}, we showed that for CP maps, the eigenvalue spectrum is always bounded. Here, our main result is that the NCP maps no longer form a compact set, essentially because the eigenvalues of the corresponding dynamical matrices are not bounded. Hence, for the set of qubit maps, including NCP maps, the volume measure may not be well defined, in general. For simplicity, we restrict ourselves to the case of unital maps. The paper is organized as follows.  In Sec. \ref{positivemaps}, we discuss the preliminaries. Section \ref{motivate} sets the idea presented in the paper through three examples. The main results are then presented  and discussed in Sec. \ref{results}. We then conclude in Sec. \ref{conclusion}.

\section{Preliminaries}
 \label{positivemaps}
\subsection{Positivity domains and valid maps}

  Let  $\mathcal{E}$ be  a positive  map acting  on a
  qubit represented by the density matrix
\begin{equation}
  \label{eq:1q1}
  \rho = \frac{1}{2}(\mathbbm{1} + a_i \sigma_i) = \frac{1}{2} \left( \begin{array}{cc}1 + a_3 & a_1 - \imath a_2\\ a_1 + \imath a_2  & 1-a_3 \end{array} \right),
\end{equation}
where the vector ${\bf  a} = (a_1 \,,\,  a_2 \,,\, a_3)$,
  with 
\begin{equation}
 a_1^2 + a_2^2 + a_3^2 \le 1,
 \label{eq:cond1}
\end{equation}  is called the  Bloch vector. All one
qubit states lie on or inside the
``Bloch   ball,''\index{Bloch   sphere}   which  is   the   unit   ball
in the space $\mathbb{R}^3$ parametrized by the axes
$a_1$, $a_2,$ and $a_3$. 

The map $\mathcal{E}$ can be represented by a four dimensional Hermitian matrix, usually referred to as the dynamical matrix~\cite{sudarshan_stochastic_1961}. The dynamical matrix is also called the Choi matrix~\cite{choi_completely_1975}. If the dynamical matrix is positive, the map is completely positive (CP). Otherwise, it is not-completely-positive (NCP). The trace of the dynamical matrix acting on a qubit is 2. 
Positivity of a map addresses its action on density matrices whereas complete positivity (CP) is a statement on the map itself. 

The important point to note is that the definition of a map makes sense only for a valid set of states. For a CP map acting on a qubit, the domain is the entire Bloch ball $\mathcal{B}$. But, for NCP maps, only certain states on the Bloch ball act as a valid domain for their action. 

\begin{Definition}[Positivity domain]
\label{Positivity domain}
Given the map $\mathcal{E}$, the set $\mathcal{P}_\mathcal{E} \equiv \{p \in \mathcal{B} | \mathcal{E}(p) \in \mathcal{B}\}$ constitutes the positivity domain of $\mathcal{E}$.
\end{Definition}
In words, $\mathcal{P}_\mathcal{E}$ is the subset of the Bloch ball $\mathcal{B}$ whose image under map $\mathcal{E}$ falls within $\mathcal{B}$.

\begin{Definition}[Valid map]
A map $\mathcal{E}$ acting on a qubit is valid iff its associated positivity domain is nonempty. i.e., $\mathcal{P}_\mathcal{E} \ne \emptyset$.
\end{Definition}
In view of this definition, a map is valid precisely if its positivity domain is nonvanishing. This definition will be motivated in the examples discussed later.

Note that the evolution of a qubit may contain a singularity at some instance, meaning that the decoherence rate in the master equation at that point, and hence the (instantaneous) intermediate map $\mathcal{E}^\ast$, diverges. The map can still be valid if the positivity domain $\mathcal{P}_{\mathcal{E}^\ast}$ is nonvanishing. That this is physically well motivated will be clarified in examples discussed later.
\subsection{Unital maps acting on a qubit}
\label{dynamicalunital}
The general form of a dynamical matrix $\tilde{B}$ acting on a qubit which is unital and trace preserving (TP) can be parametrized as follows:
\begin{equation}
\label{equnitalgeneral}
\tilde{B}= \left(
\begin{array}{cccc}
 a & x & y & z \\
 x* & 1-a & w & -y \\
 y* & w* & 1-a & -x \\
 z* & -y* & -x* & a \\
\end{array}
\right).
\end{equation}This is obtained by considering a four dimensional Hermitian matrix such that $\mathrm{tr}_{1}\tilde{B} = \mathbbm{1} = \mathrm{tr}_{2}\tilde{B}$. This means that the partial trace with respect to each of the two bipartite subsystems is identity. Note that $a$ is real and $x,y,z,$ and $w$ are in general complex. More general maps, allowing damping, can be considered, but the above suffices to prove our main result.

\section{Motivating the concept of the validity of a map}
\label{motivate}
Allowing NCP maps gives considerable freedom in what maps we can consider, even if they are not CP. Our earlier definition of validity is the only restriction we place on what we consider as valid maps. The trivial map that maps every qubit state to the fixed state $\begin{pmatrix}
2 & 0 \\ 0 & -1
\end{pmatrix}$
is invalid (as this is not a bona fide density matrix), whereas a map that takes all states to the maximally mixed state, is indeed valid, according to our definition.

\subsection{A numerical example}
\label{numex}
This numerically simulated example is to show that the eigenvalues of the dynamical matrix representing a NCP map on a qubit can exceed 2 in absolute values, unlike that for CPTP maps.
Consider a NCP map, whose dynamical matrix is given by 
\begin{equation}
\label{bnumex}
B_{NCP}=\left(
\begin{array}{cccc}
 0.20 & 0.95 & 0.70 & 0.10 \\
 0.95 & 0.80 & 0.30 & -0.70 \\
 0.70 & 0.30 & 0.80 & -0.95 \\
 0.10 & -0.70 & -0.95 & 0.20 \\
\end{array}
\right).
\end{equation}Clearly, the map is unital and trace preserving. It has eigenvalues 2.32409, -1.12409, 0.669258, 0.130742. As we can see, one eigenvalue is greater than 2, which indicates that the eigenvalues are not bounded by 2, for general NCP maps acting on a qubit. 
	Its action on a general qubit density operator
produces an output  with eigenvalues
\begin{eqnarray}
\lambda_1 &\approx& 0.5\, -0.47\sqrt{4.2 a_1^2-1.3 a_1a_3+0.04 a_2^2+2.6 {a_3}^2}, \nonumber\\
\lambda_2 &\approx& 0.5+0.47 \sqrt{4.2 {a_1}^2-1.3 {a_1} {a_3}+0.04 {a_2}^2+2.6 {a_3}^2}.
\nonumber
\end{eqnarray}
from which it follows that any choice of $a_j$ that satisfies Eq. (\ref{eq:cond1}) and 
$4.2 {a_1}^2-1.3 {a_1} {a_3}+0.04 {a_2}^2+2.6 {a_3}^2 \ge 0$ (for example, $a_3=0.5, a_1=0.05, a_2=0.1$) is an element of the positivity domain of the map. Since this is a non-vanishing set, it follows from our definition that this constitutes a valid map.
Note that in this case, the set $\mathcal{P}_\mathcal{E}$ has a non zero measure. However, this is not necessary. So long as the positivity domain is non-vanishing, the map is well defined. This is shown to be the case in the next two examples, where the positivity domain is of measure zero.

\subsection{Two successive CNOT gates \label{sec:cnot}}
Consider the action of two successive CNOT gates, $U = U_{CNOT} \circ U_{CNOT},$ on the initial product state $\frac{1}{\sqrt{2}} (|0\rangle + |1\rangle)_A \otimes |\phi\rangle_B$ of two qubits  labeled $A$ and $B$.

Let $|\psi' \rangle$ denote the total state of the systems $A$ and $B$ after the first $U_{CNOT}$ has been applied:
\begin{eqnarray}
|\psi' \rangle 
&=&  \frac{1}{\sqrt{2}} (|0\rangle_A \otimes |0\rangle_B +|1\rangle_A \otimes |1\rangle_B) ;\thinspace \text{for}\thinspace |\phi\rangle = |0\rangle, \nonumber\\
&=&  \frac{1}{\sqrt{2}} (|1\rangle_A \otimes |0\rangle_B +|0\rangle_A \otimes |1\rangle_B) ;\thinspace \text{for}\thinspace |\phi\rangle = |1\rangle.
\end{eqnarray}

Since $U = \mathbbm{1}$, the output state of $B$ is the same
as its input, which could be $|0\rangle_B$ or $|1\rangle_B$. Therefore, to discern the reduced dynamics of $B$ between the application of the two CNOT operations, one would need a map that
takes the identity state as the input, and outputs either $|0\rangle_B$ or $|1\rangle_B$. 

Note that this is a one-to-many relation and hence non-linear behavior~\cite{schmid_why_2018} would invalidate it as a map in certain conventional situations. However, this intermediate map corresponding to the application of the second CNOT is indeed valid according to our definition. This is physically well motivated since this intermediate map, only when supplemented to the CP map representing the evolution up to the application of the first CNOT, recreates the full identity map after the application of the second CNOT.

The problem may be understood as an instance of a singularity in the intermediate map, while the full map is well defined. Consider the control qubit to be in the state $\cos\theta\ket{0}+\sin\theta\ket{1}$ and the system to be the target qubit. Let us represent the map acting on the density matrix expressed as a column vector, and call it the $A$ matrix following~\cite{sudarshan_stochastic_1961,Quanta77}. Denote the $A$ matrix corresponding to the first application of CNOT to be $A_1$, and that of the second one to be $A_2$. Then we have $A_1\cdot A_2 = \mathbbm{1}_4$, from which it follows that
\begin{equation}
A_2 = (A_1)^{-1},
\label{eq:a2}
\end{equation}
meaning that $A_2$ corresponds to the inverse of the first, which is seen to be a NCP map as shown below.

The $A$ matrix for the first application of CNOT is
\begin{equation}
A_1= \left(
\begin{array}{cccc}
\cos^2(\theta) & 0 & 0 & \sin^2(\theta) \\
0 & \cos^2(\theta) & \sin^2(\theta)  & 0 \\
0 & \sin^2(\theta)  & \cos^2(\theta) & 0 \\
\sin^2(\theta)  & 0 & 0 & \cos^2(\theta)\\
\end{array}
\right),
\label{eq:a1}
\end{equation}
which corresponds to a bit flip channel.  In view of Eq. (\ref{eq:a2}), it follows from Eq. (\ref{eq:a1}) that
\begin{equation}
A_2 = \sec(2\theta)\left(
\begin{array}{cccc}
\cos ^2(\theta ) & 0 & 0 & -\sin ^2(\theta ) \\
0 & \cos ^2(\theta ) & -\sin ^2(\theta ) & 0 \\
0 & -\sin ^2(\theta ) & \cos ^2(\theta ) & 0 \\
-\sin ^2(\theta ) & 0 & 0 & \cos ^2(\theta ) \\
\end{array}
\right),
\end{equation}
from which it follows that the associated dynamical matrix is
\begin{equation}
\mathfrak{B}_2 = \sec(2\theta)\left(
\begin{array}{cccc}
\cos ^2(\theta ) & 0 & 0 &  \cos ^2(\theta ) \\
0 & -\sin ^2(\theta )& -\sin ^2(\theta ) & 0 \\
0 & -\sin ^2(\theta ) & -\sin ^2(\theta ) & 0 \\
\cos ^2(\theta )  & 0 & 0 & \cos ^2(\theta ) \\
\end{array}
\right),
\label{eq:b2}
\end{equation}
whose non vanishing eigenvalues are seen to be $-2 \sin ^2(\theta )\sec(2\theta)$ and $2\cos ^2(\theta)\sec(2\theta)$. 
Exactly one of the two eigenvalues is negative for any choice of $\theta \ne 0, n\pi$ ($n$ an integer), showing that $A_2$ is NCP for such choices. 
The parameter $\theta=\frac{\pi}{4}+n\pi$ represents a singularity, where the intermediate map represented by $A_2$ diverges, and corresponds to the nonlinearity mentioned above, which is now seen as a manifestation of the singularity rather than grounds to invalidate the map.

The operator sum representation for the map $\mathfrak{B}_2$ is 
\begin{equation}
\rho^\prime = \sqrt{1 + \sec (2 \theta )} \mathbbm{1} \rho \mathbbm{1} + \sqrt{1 - \sec (2 \theta )} \sigma_x \rho \sigma_x.
\end{equation}
Consider the fixed points of $\sigma_x$, i.e., states that are invariant under the action of $\sigma_x$. They will also be invariant under the above map, irrespective of $\theta$. Thus, when $\theta=\frac{\pi}{4}$, only these states, being invariant, will produce a valid output, whereas any other states will lead to divergent outputs. Thus, the set  of these fixed points, constitute the positivity domain.

In particular, these states have the form
\begin{equation}
\rho^X_p \equiv p\ket{+}\bra{+}+(1-p)\ket{-}\bra{-} = \begin{bmatrix}
1 & 2p-1 \\ 2p-1 & 1
\end{bmatrix},
\label{eq:X}
\end{equation}
where $0 \le p \le 1$.
Under the action of the above map, the state (\ref{eq:X}) evolves to
$[1+\sec(2\theta)]\rho^X_p + [1-\sec(2\theta)]\rho^X_p = \rho^X_p$, independently of $\theta$.
At $\theta=\pi/4$, the function $\sec(2\theta)$ and hence the map diverge, but the set $\Sigma^X_\sigma$ of all points of the form $\rho^X_p$ is unaffected by the infinity and thus constitutes our positivity domain. Notice that $\Sigma^X_\sigma$ represents a straight line connecting two extreme points on the Bloch ball, and thus has measure zero.

It is important to stress here that even when the (intermediate) map is singular-- i.e., the absolute values of the eigenvalues of the dynamical matrix increase without bound to infinity (such that their sum is 2)--the set $\mathcal{P}_{\mathcal{E}^\ast}$ is non vanishing, ensuring that the map is well-defined. This is of course validated by the fact that the full map itself corresponds to identity operation and is thus well defined.  The singularity thus gives an extreme instance of the unboundedness of eigenvalues of valid dynamical maps.

\subsection{Intermediate map with non-Markovian dephasing}
Consider an example of a non-Markovian channel based on a familiar process, namely quantum dephasing. This is described by the Lindblad equation,
\begin{equation}
\frac{d\rho}{dq} = \lambda(q) [-\rho(q) + \sigma_z\rho(q)\sigma_z],
\label{eq:lindblad}
\end{equation}
where $\lambda$ is the decoherence rate and the parameter $q$ rises monotonically from 0 to (asymptotically with time) $\frac{1}{2}$. For Markovianity to hold in the sense of CP divisibility, $\lambda$ must  always be positive as a function of  $q$ . Furthermore, for the noise to be described by a quantum dynamical semigroup, $\lambda$ must be a positive constant.

In a non-Markovian dephasing channel such as quantum random telegraph noise (RTN) \cite{daffer_depolarizing_2004}, the negativity of $\lambda$ arises from the presence of terms with (co)sinusoidal dependence on time, but in the  non-Markovian dephasing model proposed in \cite{shrikant_non-markovian_2018}
\begin{equation}
\lambda(q) = 
\frac{\frac{1}{2}(\alpha_- + \alpha_+) - q}{(q - \alpha_+)(q - \alpha_-)},
\label{eq:rate}
\end{equation}
where $\alpha_-$ is a parameter that lies in the range $[\frac{1}{2},1]$ and $\alpha_-<\alpha_+$. Thus, the regions for which $q > \alpha_-$ lead to non-Markovianity. Furthermore, $q = \alpha_-$ represents a singularity, which we discuss below. 

In Ref. \cite{shrikant_non-markovian_2018}, the authors choose
\begin{equation*}
\alpha_\pm = \frac{1}{2\nu}(\nu + 1 \pm \sqrt{\nu^2+1}),
\end{equation*}
which corresponds to the Kraus operators $K_I = \sqrt{(1-\beta)}\mathbbm{1}$ and $K_Z = \sqrt{\beta} \sigma_z$, with $\beta = [1 + \nu(1 - q)]q$, where the dimensionless quantity $\beta$ can turn negative for sufficiently large value of parameter $\nu$. In this model,  $0 \le \nu \le 1$ and serves as a non-Markovianity parameter.

 The singularity mentioned above occurring at $q = \alpha_-$ can be physically interpreted as follows. At this point, all initial states of the form
\begin{equation*}
\rho(0)\equiv\left( \begin{array}{cc} r & s \\ s^\ast & 1-r \end{array}\right), 
\end{equation*}
 are mapped to 
 \begin{equation*}
 \rho(\alpha_-) \equiv \left( \begin{array}{cc} r & 0 \\ 0 & 1-r \end{array}\right), \end{equation*}
creating an instance of non invertibility. The family of states $\rho(\alpha_-)$ corresponds to a straight line in the Bloch ball from the north to south pole, having measure zero. (For general consistency conditions pertaining to non-invertible time evolution, see~\cite{andersson_finding_2007,rivas_quantum_2014}.)

However, as in the previous example, the singularity is not pathological, and only momentary. This follows from the fact that the infinitesimal evolution determined by Eq. (\ref{eq:lindblad}), 
\begin{equation}
d\rho = \lambda(\alpha_-) [-\rho(q) + \sigma_z\rho(q)\sigma_z]dq,
\label{eq:infint}
\end{equation}
is well-behaved when acting on (only) the elements of the family $\rho(\alpha_-)$.  The intermediate map (\ref{eq:infint}) thus gives us an example of a map whose positivity domain is of zero measure. 

We shall now state and prove our main result.

\section{Results and Discussion}
\label{results}
The following result on unboundedness, which is shown for the unital maps, also can be extended to non-unital maps. But, by virtue of applying it to unital qubit maps, it obviously implies the unboundedness of the general set of maps (CP or NCP) on qubits.
\begin{Theorem} 
\label{nonbounded}
The set of maps (including NCP maps) acting on a qubit is neither closed nor bounded.  
\end{Theorem}
We represent maps in the space of eigenvalues of the dynamical matrix. By the Choi-Jamiolkowski isomorphism~\cite{jamiolkowski-1972}, the set of CP maps is isomorphic to that of two-qubit states, and hence bounded. Thus, we shall be concerned with the (un)boundedness of NCP maps. By definition, the positivity domain corresponding to any valid NCP map acting on a qubit should be non empty. Our proof proceeds by providing an example that evinces the unboundedness of  the eigenvalues of $\tilde{B}$ of Eq. (\ref{equnitalgeneral}). Specifically, consider Eq. (\ref{eq:b2}) which is in the form of Eq. (\ref{equnitalgeneral}) with $a = z = \sec(2\theta)\cos ^2(\theta ), w =  \sec(2\theta)\sin ^2(\theta ),$ and the rest are all zero. Clearly, the eigenvalues $-2 \sin ^2(\theta )\sec(2\theta)$ and $2\cos ^2(\theta)\sec(2\theta)$ diverge to $\infty$ at $\theta = \frac{\pi}{4}$, providing an instance of the unboundedness of eigenvalues of NCP maps.
This unboundedness of eigenvalues arises only for NCP maps, where the eigenvalues can take negative values. That the eigenvalues are non bounded implies that the space of NCP maps is not closed.

For NCP maps that arise as intermediate maps of a CP map, even if they diverge, one can find the positivity domain to be non vanishing, affirming the validity of the NCP map. An instance was discussed in the context of example of Sec. \ref{sec:cnot}. As another instance, consider the action of the map as in Eq. (\ref {equnitalgeneral}) on Eq. (\ref{eq:1q1}). We assume all parameters are real, for simplicity. 
The output density matrix has eigenvalues
\begin{equation}
\lambda_{\pm} = \frac{1}{2}\Big(1\pm\sqrt{a_1^2 \left((w+z)^2+4 x^2\right)+4  a_1a_3 ((2 a-1) x+y (w+z))+a_2^2 (w-z)^2+a_3^2 \left((1-2 a)^2+4 y^2\right)}\Big).
\label{eq:ev}
\end{equation}
In this case, setting the positivity condition on the two eigenvalues, one can see that the point $(a_1,a_2,a_3) = (0,0,0)$ corresponding to the maximally mixed state serves as one point in the positivity domain of the NCP map. In general, one can find sets of points $(a_1,a_2,a_3)$ in the Bloch ball such that the eigenvalues $\lambda_{\pm}$ in Eq. (\ref{eq:ev}) are both positive. This set will constitute the positivity domain of the map (for an illustration, see below).
\hfill $\blacksquare$ 
\bigskip

\begin{figure}[t!]
\resizebox{6 cm}{6 cm}{\includegraphics{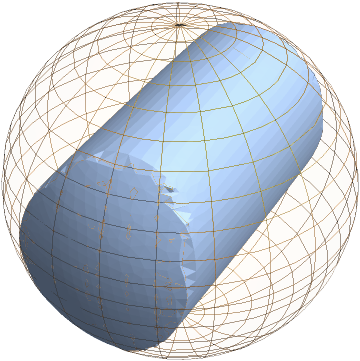}}
\caption{(Color online) The positivity domain for the NCP map represented by the dynamical matrix in Eq. (\ref{bnumex}). The outer ball (of unit radius) shown in white is the Bloch ball. \label{fig1}} 
\end{figure}

For the case of Eq. (\ref{bnumex}) in the example discussed in Sec. \ref{numex}, the positivity domain is plotted in Fig. (\ref{fig1}). That this domain is a subset of the Bloch ball implies that the map in question is not positive, and thereby goes beyond the restriction under which a simple measure of NCP maps was found in Ref. \cite{jagadish_measure_2019}.
This result for the unital map can be extended (albeit in more cumbersome way) to the non unital case, also. However, the present example suffices for the proof.

It follows from the unboundedness of NCP maps, that the set of qubit maps, including NCP maps, do not form a compact set. Now, if the set of divergent maps form a discrete set of singularities, then one might hope that the set of maps obtained by excepting these points could be endowed with a bona fide measure. However, it is easy to see that the divergent maps themselves form a continuous family.

To see this,  note that we can replace the CNOT in the double-CNOT example with any other entangling operation, and we obtain a divergent map corresponding to the intermediate map represented by the second application of the control operation. Indeed, consider any control-$Q(\theta,\phi, \xi)$, where $Q(\theta,\phi, \xi)$ is any unitary, i.e.,
$$
I_2 \otimes I_2 + I_2 \otimes Q(\theta,\phi, \xi),
$$
and the angles $\theta, \phi,$ and $\xi$ may be varied continuously. This strongly suggests that a well-defined measure of maps for this general case does not seem to arise, unless further restrictions are made.

We enumerate a few such interesting cases:
\begin{enumerate}
	\item If we restrict to the case of Pauli channels, subject to the condition of positivity (i.e., having full positivity domain), then the volume measure of NCP maps is twice that of CP maps \cite{jagadish_measure_2019}.
	\item A similar result also holds if the Pauli channels are rotated by a fixed unitary $U$. This corresponds to the simplex whose four vertices are $UPU^\dagger$, where $P$ is a Pauli matrix. This may be considered as a rotated version of the Pauli tetrahedron $\mathcal{T}$, discussed in \cite{jagadish_measure_2019}.
\end{enumerate}
Finally, note that the double-CNOT, or equivalently a double-C-phase gate, provides perhaps the simplest method to practically realize NCP maps, and even  a singularity, conveniently on a quantum computer in a way that is well within the reach of present-day quantum technology. Indeed, the double-control application of any two-qubit gate will be acceptable, based on the argument given above.

\section{Conclusions}
\label{conclusion}

Positivity domains for NCP qubit maps, arising from initial correlations with an environment, were first addressed in~\cite{jordan_dynamics_2004}, where it was shown that they can be strict subsets of the Bloch ball. Here, we show that arbitrary (even divergent) NCP maps, subject only to the trace-preserving condition, can be physically valid in the sense of having a non-vanishing (possibly zero-measure) positivity domain.

The physical significance of this result is that the volume measure of qubit maps, including NCP maps, may in general not be well defined, even though the dynamics is quite regular~\cite{chrusinski2010nonmarkovian}. Specific restrictions may be imposed in certain cases, as in~\cite{jagadish_measure_2019}, to define a measure.

\section{Acknowledgements}
The work  of V.J.  and F.P.  is based upon  research supported  by the
South African Research  Chair Initiative of the  Department of Science
and  Technology and  National Research  Foundation.  R.S.   thanks the
Defense Research  and Development  Organization (DRDO), India  for the
support      provided       through      the       Project No. ERIP/ER/991015511/M/01/1692.
\bibliography{Affine}

\end{document}